# Towards a New Participatory Approach for Designing Artificial Intelligence and Data-Driven Technologies

Soaad Qahhār Hossain
University of Toronto
Toronto, ON, Canada
soaad.hossain@mail.utoronto.ca

Syed Ishtiaque Ahmed
University of Toronto
Toronto, ON, Canada
ishtiaque@cs.toronto.edu

## ABSTRACT

With there being many technical and ethical issues with artificial intelligence (AI) that involve marginalized communities, there is a growing interest for design methods used with marginalized people that may be transferable to the design of AI technologies. Participatory design (PD) is a design method that is often used with marginalized communities for the design of social development, policy, IT and other matters and solutions. However, there are issues with the current PD, raising concerns when it is applied to the design of technologies, including AI technologies. This paper argues for the use of PD for the design of AI technologies, and introduces and proposes a new PD, which we call *agile participatory design* that not only can could be used for the design of AI and data-driven technologies, but also overcomes issues surrounding current PD and its use in the design of such technologies.

**ACM Reference Format:**
Soaad Qahhār Hossain and Syed Ishtiaque Ahmed. . Towards a New Participatory Approach for Designing Artificial Intelligence and Data-Driven Technologies. In . ACM, New York, NY, USA, 5 pages.

## 1 INTRODUCTION

Marginalized communities is a group of individuals that are excluded from mainstream social, economic, educational, and/or cultural life [8]. Within the research and development of artificial intelligence (AI) technologies, a growing concern that these technologies will not have marginalized community members in mind. That is, the AI technologies will be designed and developed for those that mainstream social, economic, educational, and/or cultural life. Consequently, the use of AI technologies can negatively impact marginalized communities. To address this, we argue for the use of participatory design (PD), an approach widely used with marginalized people, for the design of AI technologies. In this paper, we will first review related work, followed by a discussion on the use of PD for the design AI technologies. Then, we will propose and elaborate on a new PD that can be used for the design of AI technologies. We will conclude with addressing challenges associated with AI and PD, and a discussion on the new PD with respect to the design of AI technologies and data-driven technologies.

Our work contributes to the area of human-computer interaction (HCI) and human-centered computing (HCC) as it provides a design method used with marginalized people that may be transferable to the design of AI technologies. Furthermore, our work contributes to the area of AI and machine learning it expands on the discussion of the role of PD in the research and development of AI and machine learning, and provides a PD that could be used for the design of AI technologies and data-drive technologies.

## 2 RELATED WORK

### 2.1 Participatory Design and Research, Marginalized Communities and HCI

PD is known as a a design method in which users and other stakeholders work with designers in the design process [7]. Figure 1. displays a model of PD often used for design and research. From a more technical standpoint, PD can be considered as an approach that aims to design systems while building a connection and promoting active work between its targeted population group and its developers [4]. Technical designers, IT professionals and HCI researchers consider PD due to its high potential for enabling users to be active contributors during the development of computer-related products and activities [4]. Additionally, PD is frequently advocated when developing new solutions for economically or socially marginalised people in developing countries [7]. In addition, there are several studies that engage with PD in research and development of solutions. In Byrne and Sahay's case study, they investigated the role of participation within the domains of information system (IS) research and social development in South Africa, and found that PD needs to be more clearly defined and re-articulated in the context of contemporary IS research. In addition, they provided three key aspects of participatory design in developing countries: (1) different strategies of participation; (2) scaling of participation; and (3) contextual specificity [2]. In Teeter's case study, her work demonstrated ethnography and PD are necessary to generating equity-oriented design solutions [11].

An HCI study by Duarte et al. reported a deeper understanding of advantages and disadvantages of using PD and participatory research (PR) with young forced migrants. The main advantages were creating a safe space for young forced migrants participation; possibility of empowerment and engagement because of the activities; and more participation and impact through involving young forced migrants [4]. The main disadvantages were challenges associated with engaging with young forced migrants of the host community; specifically limited common language proficiency, engaging with the young forced migrants in a formal education setting in the participatory activities, and compelling communication of collected data management procedures to young migrant workers [4]. An HCI study by Bødker and Kyng found that while PD is helpful for designing technological solutions due to it empowering of users and their role as design partners, PD needs to be redesigned in order for it to be applicable for current technologies and issues that matter [3]. To elaborate, they state that most present-day PD is on how to facilitate direct collaboration between users and designers in codesign processes to engage with everyday issues of use, through technology or otherwise; consequently, with that the role getting adapted by HCI at large, this led to what we see as a focus on small issues (in contrast to big and important ones) such as products and technological solutions that the users like, rather than on solutions that profoundly change their activities as well as the goals they are supported in pursuing [3]. Accordingly, within their work, they discuss and advocate for the possibilities for a new PD - one that explores larger scales, develop strong alliances, address changes that matter, be based on engaged partners, have a vision for high and lasting impact, strives for democratic control of information technology (IT), and have PD researchers that are also activists [3].

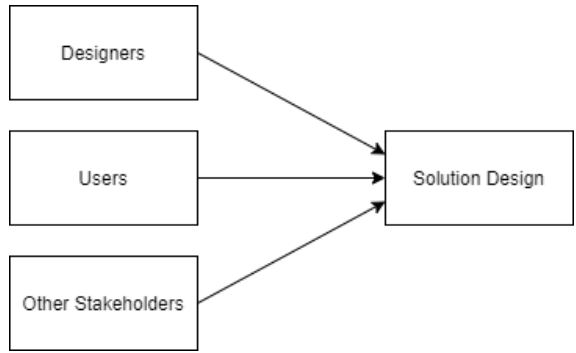

**Figure 1: Traditional model for participatory design.**

## 2.2 Participatory Design and AI

There are are only two piece of literature that covered PD and AI. A work that directly addressed participatory design in AI technology was by Bratteteig and Verne. In their article titled "Does AI make PD obsolete? Exploring challenges from Artificial Intelligence to Participatory Design", they argued that AI does make PD obsolete. The reasons they provided, which they also frame them as challenges for PD in the design of AI technologies, is that it is difficult for the technical aspects of AI to be explained to users and stakeholders; that because AI technologies constantly change in the sense that they are always learning, to involve users and stakeholders would not productive; and that the involvement of users and stakeholders can limit the development of the AI technology as it is difficult to consider or foresee mathematical and statistical approach and the training data set sample in an early state of the design process [1]. Note that when they are talking about AI technologies, they are only talking about AI technologies that rely on neural networks as their learning algorithm. We will further discuss and address Bratteteig and Verne's challenges later in the paper. Excluding the challenges, in their work, they suggested making use of more elaborate design fiction or scenarios that can be played out to help with the designing AI [1].

A recent work by Gyldenkærne et al. called for a PD approach to understand and reconfigure the socio-technical setup in healthcare especially where AI is being used on EHR data that are manually being submitted by healthcare professionals. The idea of using a PD approach comes from a study of AI being used to predict patient no-show's, which led them to realize that for the AI to gain full potential there lies a need to balance the introduction of AI with a proper focus on the patients and the clinicians' interests [5]. In their work, Gyldenkærne et al. found that the role of PD is to ensure the voice of the user and end-user; that is, balance the introduction of AI by maintaining a proper focus on the patients and the clinicians closely involved in their treatment and care [5]. Furthermore, their work concludes with stating that AI introduces an occasion PD might contribute by designing re-configurations of work and technologies that benefit both types of use and purpose of EHR data [5].

## 3 USE OF PARTICIPATORY DESIGN FOR THE DESIGN OF AI TECHNOLOGIES

Literature in HCI and elsewhere provide substantial reasons for using PD for technical solutions, with the main and overlapping reason being that using PD enables for solutions to be designed in a way that takes into consideration the voices of its users and other stakeholders. For the same reason, we also support and use PD, but for the design of AI technologies. In using PD for the design of AI technologies, we can approach the design of AI technologies in a way that truly includes the voices of marginalized communities. Furthermore, in using PD for the design of AI technologies, this also creates opportunities to building connections and promoting active work between those designing and developing AI technologies and those within marginalized communities, and create dialogues for how AI technologies can be better be utilized by marginalized communities and address ethical and other concerns pertaining to AI. Ultimately, PD enables for marginalized communities to help designers and developers with attaining equity-oriented design of AI technologies and increase the level of trust of marginalized communities and AI, and even machine learning, technologies.

We know that the main aim of participatory processes in social development is to involve marginalized people meaningfully in the decisions that affect their lives [2]. However, we also know that current participatory processes in technological development focuses on involving people, which those people are not necessarily marginalized people, but in small issues such as products and technological solutions that the users like, rather than big issues such has decisions that affect their lives and other people's lives [3]. Consequently, to apply current PD to the design of AI technologies will likely run into several major issues, many which have been expressed within HCI literature. To overcome those issues and other issues that matter, such as the challenges from AI to PD, we need to create a new PD.

## 3.1 Design of AI Systems

The way that AI systems, and many contemporary data-driven systems are approached is essentially the same as how predictive analytics tools are approached. There are four phases to predictive analytics projects [6]:

- Phase 1: Objectives – Phase where objectives are defined.
- Phase 2: Data Collection – Phase where data is acquired.
- Phase 3: Model Development – Phase where a mathematical model is built.
- Phase 4: Model Application – Phase where the mathematical model is used in practical settings.

The mathematical model would either be a classification model or a predictive model. With AI projects, as AI generally deals with big data, there may be some additional steps depending on the data set being used. Additional phases include a phase for data cleansing phase and a phase for dimensionality-reduction, which one or both of those phases can take place after the data is collected and before the final mathematical model is built.

## 3.2 New Participatory Design for AI Technologies

To approach the new PD, which we will call *agile participatory design*, we will break it down into three sections: participants, alliances, and involvement.

*3.2.1 Agile Participatory Design Project Participants.* A portion of the users and stakeholders will be activists for a marginalized communities and/or marginalized community members. The reason why we used "and/or" instead of "and" is because it is not always possible for marginalized community members to get involved in projects due to infrastructural, language and other barriers that prevent them from participating. In having activists involved in their place, they can advocate for the needs of marginalized communities. Note that activists are not simply just people that claim that they support at least one social cause - they are people that are trusted and known within the community for their involvement in social activities such as protests, marches, and other forms of activism. In addition to the activists, among the stakeholders, a portion of them should be engaged partners. This is essentially realizing one of Bødker and Kyng's ideas within their work, which is the engaged partners being users that are closely related to the changes they experience or expect, and to their hope for major, lasting changes in areas that are important to them. In terms of who these engaged partners would be, they would be ethicists, public health experts, public interest legal practitioners, social workers, and others that care for the well-being of individuals and the public. In having all of these people as users and stakeholders, this can help with having a PD for the design of AI technologies to focus on larger issues, changes that matter, have a vision for high and lasting impact, and democratic control of AI technologies.

*3.2.2 Agile Participatory Design Project Alliances.* One problem with PD stems from the fact that they are usually limited to research papers [3]. Consequently, for AI technologies where the AI system extends beyond a research paper or report, it may become a challenge to continue working for realizing the vision agreed upon by the designers, users and other stakeholders, including getting support for deployment of AI system developed. To overcome this, alliances need to get formed. These alliances would be between the management and users and stakeholders. In creating such alliances, this can help ensure that the AI system deployed will be one that met the needs of all users and stakeholders, including those from marginalized communities and/or activists, allowing for the high and lasting positive impact to be successfully realized. These alliances will also help users and stakeholders with fighting off negative changes proposed to the design of the AI technology.

*3.2.3 Agile Participatory Design Project Involvement.* The way how agile PD came about was from the realization of the overlaps between agile software development and PD, and the issues with present-day PD. Like in agile software development, users and stakeholders will be involved several phases within the AI system. Similarly, just as for agile software development, success is highly dependent on how the software solution fulfils the expectations of the users addressed [10], agile PD makes the design of AI technology success becomes highly dependent on how the AI fulfils the expectations of the users and stakeholders. Figure 2. illustrates the agile PD for the design of AI systems. For confidentiality reasons, the dataset(s) must be anatomized when shown to the users and other stakeholders. Only key information, such as race and ethnicity, will be shared with them. Prior to the final phase, only the management that allied with the users and others stakeholders will be able to approve the machine learning model for deployment.

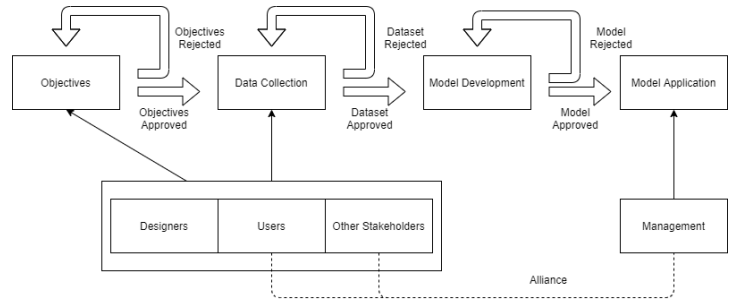

**Figure 2: Model for agile participatory design in the design of AI systems.**

## 4 CHALLENGES BETWEEN AI TECHNOLOGIES AND PARTICIPATORY DESIGN

In Bratteteig and Verne's work, they argued that AI technologies pose three challenges to participatory design. In this section, we will acknowledge the challenges, and address them accordingly. The reasons why we are addressing these challenges is because these challenges are not just concerns from them, but they are concerns that are shared by many people within the AI community, and other communities involved in data-driven applications.

## 4.1 Machine Learning Comprehension

The first challenge stated in Bratteteig and Verne's work involves solely the designers. Just as with any sort of technology, designers should be able to understand the technology they make. However, in the case of AI technologies, designers cannot foresee the effects of their designs due to the hidden deep inside the statistical machinery and in the training data set [1]. Consequently, if they themselves do not know understand how the AI technology makes the decision that it made, then it will difficult for them to explain it to users and other stakeholders.

The first thing that we need to clarify is how Bratteteig and Verne define machine learning within their work. They defined it exactly as follows: "ML is concerned with training a neural network to recognize particular patterns or properties in the data material.", which ML stands for machine learning. This however is false. Machine learning is not concerned with training a neural network; neural network is a type of learning algorithm. Machine learning is a subset of artificial intelligence, which build a mathematical model based on sample data (i.e. training data) in order to make predictions or decisions without being explicitly programmed to perform the task [12]. Accordingly, machine learning is concerned with building a mathematical model that accurately makes predictions or decisions, and not with training a neural network, specifically. The mathematical model can be a neural network, but not all mathematical models in machine learning are neural networks. As such, while what Bratteteig and Verne said may be true in the case where the AI technology relies on a neural network (due to the explainability issue associated with neural networks), but it is not true in other cases. An AI technology that utilizes k- Nearest Neighbors (kNN) to make predictions or decisions, is an example where the first challenge stated by Bratteteig and Verne does not hold. Designers can easily explain decisions or predictions made by kNN to a user or stakeholder. However, in the case where an AI technology utilizes support vector machine (SVM), the first challenge brought forth by Bratteteig and Verne applies as for SVMs, it is difficult to explain their predictions due to implicit mapping done in kernel classification is uninformative about the position of data points in the feature space and the nature of the separating hyperplane in the original space [9]. Nonetheless, the challenge is not something that is guaranteed to occur for every AI technology, making the challenge less of an issue than what Bratteteig and Verne made it to be.

## 4.2 Participatory Design and Development

The second challenge stated in Bratteteig and Verne's work is that for the participatory design project participants to evaluate a design idea within a participatory design project time frame; it is only possible to evaluate the design result over time. They also said that evaluation as a step in design aimed at judging if a design decision is right loses its meaning, and that instead, we need to discuss possible futures as a basis for arriving at designs that avoid unwanted outcomes [1]. We do agree with them on the idea of discussing possible futures as a basis for arriving at designs that avoid unwanted outcomes, especially for AI technologies relying on deep learning or learning algorithms with explainability issues such as SVM. One of the possible futures is the that of the marginalized communities and them being negatively impacted by AI technologies, which is something that the HCI and other communities are discussing about. Our work contributes to this discussion as it provides a design that could be used to avoid unwanted effects from AI technologies to marginalized communities. As for the part concerning participatory design project participants and time frame, in using agile PD, that challenge gets addressed through the alliances made between the management, users and stakeholders.

## 4.3 AI System Training and Application

The third challenge stated in Bratteteig and Verne's work is concerned with how to distinguish between normal use and training; an AI is trained while being used: use is training. They continue with saying how that additional dimension of use may make the user too aware of the technology to not teach it the wrong things, which that may limit the value of the AI technology for the user [1]. The issue with this challenge is that because the AI technologies currently obtain most of their (training) data from non-marginalized communities and generally excludes marginalized community members' voices, regardless of whether the additional dimension of use is there or not, the value of the AI technology for the marginalized community users will be limited. There can be moments where the AI technology can be useless or even detrimental for the marginalized community member(s). A possible way to address this issue is through getting marginalized community members involved in the designing process of the AI technology, which is essentially applying participatory design to the design of AI technology. In getting them involved, they can help ensure that designers are using data sets that are diverse and contain individuals from marginalized communities, and that things such as machine learning fairness is enforced within the AI technology.

## 5 CONCLUSION

PD can be transferable to the design of AI technologies. Agile PD is not the perfect solution to the problems associated with the design of AI and data-driven technologies. There are likely things that we may have excluded when describing agile PD that should have been incorporated. An example would be having users and other stakeholders involved with the machine learning algorithm selection; some may argue that they should be involved in the algorithm or model selection process. However, agile PD is a first major step towards having a design method used with marginalized people that may be transferable to the design of AI technologies, but also revamped so that it does not encounter and contain the issues that exist with present-day PD. The next step would be to further investigate, modify and expand on agile PD so that it can cover issues that are important but were not brought up in this paper, then testing agile PD to obtain empirical results needed to validate the design method. Expanding on agile PD then testing it will allow it to become a design method that works with marginalized people and for the design of AI and data-driven technologies.